\documentclass{article}

\usepackage{amssymb,amsfonts,amsmath,stmaryrd}
\usepackage{cite,enumerate,float,indentfirst}
\usepackage{color}

\def\be{\begin{eqnarray}}
\def\ee{\end{eqnarray}}
\def\nn{\nonumber}

\def\p{\partial}

\def\Tr{{\rm Tr}\,}

\def\Ker{{\rm Ker}}

\definecolor{red}{rgb}{1,0,0}
\definecolor{orange}{rgb}{1,0.5,0}
\definecolor{violet}{rgb}{0.7,0,1}

\hoffset= - 1.0in         

\begin{document}

\title{\vspace{-.cm}{\Large {\bf  On Hamiltonians for Kerov functions}\vspace{.05cm}}
\author{
{\bf A.Mironov$^{a,b,c}$}\footnote{mironov@lpi.ru; mironov@itep.ru}\ \ and
\ {\bf A.Morozov$^{d,b,c}$}\thanks{morozov@itep.ru}}
\date{ }
}

\maketitle

\vspace{-5cm}

\begin{center}
\hfill FIAN/TD-09/19\\
\hfill IITP/TH-12/19\\
\hfill ITEP/TH-20/19\\
\hfill MIPT-TH-10/19
\end{center}

\vspace{3cm}

\begin{center}
$^a$ {\small {\it Lebedev Physics Institute, Moscow 119991, Russia}}\\
$^b$ {\small {\it ITEP, Moscow 117218, Russia}}\\
$^c$ {\small {\it Institute for Information Transmission Problems, Moscow 127994, Russia}}\\
$^d$ {\small {\it MIPT, Dolgoprudny, 141701, Russia}}
\end{center}

\vspace{.5cm}

\begin{abstract}
Kerov Hamiltonians are defined as
a set of commuting operators
which have Kerov functions as common eigenfunctions.
In the particular case of Macdonald polynomials, well known are
the exponential Ruijsenaars Hamiltonians,
but the exponential shape is not preserved in lifting to the Kerov level.
Straightforwardly lifted  is a bilinear expansion
in Schur polynomials, the expansion coefficients being factorized
and restricted to single-hook diagrams.
However, beyond the Macdonald locus, the coefficients do not celebrate these properties, even for the
simplest Hamiltonian in the set.
The coefficients are easily expressed in terms of the eigenvalues,
and, if so defined, the set of commuting Hamiltonians
is enormously large: one can build one for each arbitrary set of
eigenvalues $\{E_R\}$, specified independently for each
Young diagrams $R$.
A problem with these Hamiltonians is that they are constructed with the help of Kostka matrix instead of defining it, and thus are less powerful
than the Ruijsenaars ones.
\end{abstract}

\vspace{1cm}

{\bf 1.} Symmetric polynomials like Schur and Macdonald polynomials
play an increasingly important role in string theory studies.
Modern theory of 6d models and AGT relations \cite{AGT} is fully formulated
in these terms \cite{AF,SF1,genmac,japsgenmac}, as does \cite{SF2,MMMI} its emerging extension to Chern-Simons theory \cite{CS}
and knot polynomials \cite{knotpols}.
This adds to the prominent role these polynomials played in
description of integrable structures, i.e. of the properties of
generic non-perturbative partition functions.
At the same time, the theory of Macdonald polynomials {\it per se} \cite{Mac}
remains somewhat fragmentary and is still more a piece of art than
a solid construction from the first principles.
A part of the problem here is an emphasis on similarity with
more simple Schur polynomials, which are simultaneously characters of linear
groups $GL_N$ and thus have a clear representation theory interpretation.
The Macdonald polynomials preserve many of these connections to
representation theory, but not all.
Moreover, the simplest algebra to which they are truly related \cite{AF,AFS,DIMMac,AR}
is the enormously big DIM algebra \cite{DIM},
however they are not generic in this framework,
but instead occupy just a small corner in the set of
still under-investigated MacMahon characters and 3-Schur functions \cite{MacMahon}.
Thus the true group theory meaning of Macdonald polynomials remains obscure,
but concentration on the representation properties
overshadows other aspects of their story,
which can finally be a big mistake.

From this point of view, it is important that Macdonald polynomials
possess a generalization not only in the MacMahon (DIM) direction,
but also in a seemingly different one,
to the Kerov functions \cite{Kerov}
(see \cite{FZ,Cha} for early  applications and \cite{MMkerov}
for a recent review).
The Kerov functions break direct links to representation theory
and leave only those to Young diagrams:
multiplication of the Kerov functions does not respect peculiar representation theory zeroes, e.g.
\be
\widetilde\Ker_{[4]}\cdot \widetilde\Ker_{[1,1]} =
\tilde\alpha\,\widetilde\Ker_{[5,1]} + \tilde\beta\,\widetilde\Ker_{[4,2]}
+\tilde\gamma\,\widetilde\Ker_{[3,3]}+\tilde\delta\,\widetilde\Ker_{[4,1,1]}
\label{KLRvio}
\ee
The two Kerov-Littlewood-Richardson numbers $\tilde\beta$ and $\tilde\gamma$
vanish only at the Macdonald locus, since the representations of $GL_N$ associated with
$[4,2]$ and $[3,3]$  do not appear in the product of $[4]$ and $[1,1]$.
However, as symmetric polynomials, the Kerov functions are the most natural objects,
defined as a natural deformation of Schur polynomials
induced by a minor change of the scalar product in the space of time-dependent
functions\footnote{
Throughout this paper, we use the standard group/knot theory notation:
for the Young diagram
$\Delta = \left[\delta_1\geq\delta_2\geq \ldots \geq \delta_{l_\Delta}>0\right]
=  \big[\ldots ,\underbrace{2,\ldots,2}_{m_2}, \underbrace{1,\ldots, 1}_{m_1}\big]$,
the size (level) is $|\Delta| = \sum_{i=1}^{l_\Delta} \delta_i = \sum_a am_a$,
the time-monomial is $p_\Delta := \prod_{i=1}^{l_\Delta} p_{\delta_i}$,
and combinatorial factor is $z_\Delta = \prod_{a} a^{m_a} m_a! $.
Also, $\{x\} :=x-x^{-1}$. We deal with the Schur, Macdonald polynomials and the Kerov functions, which are symmetric polynomials of variables $x_i$ as functions of
time variables $p_k:=\sum_ix_i^k$, they are labeled by Young diagrams, we use the
notation $\chi_R\{p\} = {\rm Schur}_R\{p\}$ for the Schur polynomials.
}
\be
\Big<p_\Delta \Big| p_{\Delta'}\Big> \ = \ z_\Delta\cdot\delta_{\Delta',\Delta}\cdot g_\Delta
\label{pprod}
\ee
from $g_k=1$ to $g_k\neq 1$.
Moreover, considering $g_k$ as a new set of time variables, the Macdonald locus can be treated just as a counterpart of
the topological locus (a special point in the space of time variables, where the Schur polynomials reduce to the quantum dimensions \cite{MMMI}: $p_k=\{q^{Nk}\}/\{q^k\}$), only this time it is in the space of Kerov times,
\be
g_k^{\rm Mac} = \frac{\{q^k\}}{\{t^k\}}
\label{Macloc}
\ee
Clearly, the study of Kerov functions should provide a new dimension to
understanding of the Macdonald polynomials, and, given the now-undisputable
significance of the latter, this is already a reason.
However, there is little doubt that one day significance of the Kerov functions
will grow much further.
The very first attempts \cite{MMkerov} demonstrate that they are infinitely
more sophisticated and richer than the Macdonald polynomials, and, at the same time,
possess just the same properties and can be handled by just the same methods.
This feature , direct generalization, which preserves known properties and
technical tools, but provides considerably heavier answers
is a standard sign of ``new physics", which promises great insights in the
application to the old subjects (representation theory, knots, integrability,
non-perturbative calculations) and gives hopes to new applications in some
unpredictable directions.

Our main concern in this paper will be the long-standing problem of {\bf Kerov Hamiltonians},
which we do not truly resolve, but at least explain what can be achieved
easily, and what can not.
Accordingly, the presentation is split into three parts.
In secs.{\bf 2}--{\bf 6}, we remind some known facts about the Kerov functions
and the Macdonald Hamiltonians, putting them in the form which we need for our purposes.
Then in secs.{\bf 7}--{\bf 9}, we elaborate on the particular realization
of the naive Hamiltonians (\ref{obviousHam}) in the Kerov case.
Finally in secs.{\bf 10}--{\bf 12}, we discuss the options and obstacles
for construction of truly interesting Hamiltonians, which can play in the Kerov case
the same role as peculiar exponential Ruijsenaars Hamiltonians play in the
particular case of Macdonald polynomials.
Sec.{\bf 13} is a brief conclusion, summarizing what we could and could not achieve so far. The Appendix contains useful formulas for calculating using hook diagram Schur polynomials and some illustrative examples to the main body of the text.

\bigskip

{\bf 2.} There are two rather different approaches to the {\it definition}
of Macdonald polynomials: they are
\begin{itemize}
\item[{\bf (a)}]  triangular linear combinations of the Schur polynomials with respect to the lexicographic ordering of Young diagrams that are obtained by orthogonalization procedure with respect to the scalar product (\ref{pprod})+(\ref{Macloc}), and
\item[{\bf (b)}] common eigenfunctions of the Ruijsenaars exponential Hamiltonians \cite{RHam}
$\hat H_m$, the simplest of which is In fact, just this Hamiltonian is enough to fix the Macdonald polynomials unambiguously)
\be
\hat H_1 =
 \oint\frac{dz}{z}
 \exp\left(\sum_{k>0}{(1-t^{-2k})\,p_kz^k\over k}\right)\cdot
 \exp\left(\sum_{k>0}
 {q^{2k}-1\over z^k}
{\partial\over\partial p_k}\right)
\label{RHam1}
\ee
\be
 \frac{\hat H_1-1}{t^2-1}\  {\rm Mac}_{_R}\{p\}
= \left(\sum_{i=1}^{l_R} \frac{q^{2r_i}-1}{t^{2i}}\right)
\cdot {\rm Mac}_{_R}\{p\}
\label{evRMac}
\ee
(see \cite{AR,MMgenmac} and sec.{\bf 6} below for higher Hamiltonians).
\end{itemize}
The triangularity is not immediately obvious from these Hamiltonians,
at the same time, it is the triangularity (orthogonalization procedure),
which provides the most efficient way to calculate.
On the other hand, at least one further generalization is known:
to {\it generalized} functions \cite{genmac},
where the triangularity (definition {\bf (a)}) is still not enough to provide the answers
\cite{MMgenmac},
only {\it a generalization} of the Hamiltonians (definition {\bf (b)}) works
\cite{japsgenmac}.
Thus, at least to approach the issue of {\it generalized} Kerov polynomials,
one needs Kerov Hamiltonians, i.e. a deformation of (\ref{RHam1}) to
arbitrary $g_k$.
This is a {\it need}, but, of course, understanding the Hamiltonians
is a necessary step to make in the course of studying the Kerov functions,
irrespective of any particular needs or applications.

Expected or not, but the exponential shape (\ref{RHam1})
is violated by the Kerov deformation.
To find Kerov Hamiltonians, one needs another approach.
Moreover, in order to get a perspective, it is better to return to the level of
Schur polynomials, where we will find three different approaches to
the problem, and one of them will allow a direct lifting to the Kerov case.

\bigskip

{\bf 3.}
For the Schur polynomials, the most natural is a set
of commuting cut-and-join operators $\hat W_\Delta$ \cite{MMN1},
for which the Schur polynomials $\chi_R$ are the common eigenfunctions:
\be
\hat W_\Delta \chi_R = \psi_R(\Delta) \chi_R
\ee
The simplest non-trivial of these operators is the celebrated cut-and-join operator \cite{GJ}
\be
\hat W_{[2]} = \frac{1}{2}  \sum_{a,b} \left((a+b)p_ap_b\frac{\p}{\p p_{a_b}}
+ abp_{a+b}\frac{\p^2}{\p p_a\p p_b}\right)
\ee
hence the name for the entire family.
Eigenvalues $\psi_R(\Delta)$ are also interesting: they are characters of the symmetric groups, with orthogonality properties
\be
\sum_R  \psi_R(\Delta)\psi_R(\Delta') = z_\Delta \delta_{\Delta,\Delta'}  \nn \\
\sum_\Delta \frac{\psi_R(\Delta)\psi_{R'}(\Delta)}{z_\Delta} =  \delta_{R,R'}
\ee
and the Fr\"obenius formula
\be
\chi_R\{p\} = \sum_\Delta \frac{\psi_R(\Delta)}{z_\Delta} \cdot p_\Delta
\ee
at $|R|=\Delta|$, and are naturally continued to $|R|>\Delta|$ by adding the necessary number of the unit cycles to $\Delta$ (see details in \cite{KI,MMN1,MMNspin}).

The operators $\hat W_\Delta$ form a commutative ring with interesting
(and still not fully known) structure constants,
where $W_{[m]}$ for symmetric representations $R=[m]$ form a multiplicative
basis, i.e. $W_\Delta = \ :\prod_{i=1}^{l_\Delta} W_{[\delta_i]}: \ +\ {\rm corrections}$.

Unfortunately, the $q,t$-deformation of cut-and-join operators $\hat W$
even to the Macdonald polynomials
is under investigated, despite these operators can seem closely related to $GL_N$.
Indeed, the simplest realization of
$\hat W_\Delta$ is by the matrix derivative operator
$\ :\prod_{i=1}^{l_\Delta} \Tr \left(X\frac{\p}{\p X}\right)^{\delta_i}: \ $ (the normal ordering here means pushing all derivatives to the right), where the matrix $X$ is related to the time variables via $p_k = \Tr X^k$, and there is no direct way
to extend {\it this} definition to the Macdonald case, nothing to say about the Kerov one.
Still, the Macdonald deformation seems to exist, but has not yet been worked out,
see \cite{MmacW} for a preliminary description.
The question about Kerov deformation remains open.

Deformable to the Macdonald case is the operator (\ref{RHam1}) with $q=t$,
but $t$-dependence still survives (!).
Thus for the Schur polynomials, which are independent of $q$ and $t$,
this is a whole {\it family} of operators, which is sufficient to define all of
the Schur polynomials, no higher Hamiltonians are needed.

More important, {\it an origin} of (\ref{RHam1}) remains obscure,
including the reasons why it has such a spectacular simple exponential shape
equivalent to describing it as a shift operator,
which makes the Macdonald polynomials out of solutions of the {\it difference} equations.
It is at best {\it unclear}, if one can expect any difference equations for the Kerov functions.
As we shall see below in this paper,
the shape (\ref{RHam1}) implies some kind of {\it factorization},
which is violated by the Kerov deformation,
and this can explain the failure of attempts to generalize (\ref{RHam1})
to the Kerov functions directly.

\bigskip

{\bf 4.}
Fortunately, at the Schur level, there is still a third approach,
originally discussed in \cite{MMsumrules} and recently, once again, in \cite{triangle}.
Namely, one can represent {\it a} Hamiltonian as a bilinear combination of the Schur polynomials:
\be
\boxed{
\hat{\cal H} = \sum_{X,Y} \xi_{_{X,Y}}\chi_{_X} \,\hat\chi_{_Y}
\ \ \ \  \Longleftrightarrow \ \ \ \
\hat{\cal H}\,\chi_{_R}\{p\} =  \sum_{X,Y} \xi_{_{X,Y}}\chi_{_X}\{p\}\,\chi_{_{R/Y}}\{p\}
}
\label{bilH}
\ee
Here we use the standard notation $\hat\chi_Y = \chi_{_Y}\!\!\left\{k\frac{\p}{\p p_k}\right\}$
for  the Schur polynomial depending on time derivatives instead time variables,
and the fact is that it acts on the Schur polynomials converting them into the skew Schur ones:
\be\label{diff1}
\hat\chi_{_Y}\chi_{_R}\{p\} = \chi_{_{R/Y}}\{p\}
\ee
Putting all $p_k=0$ in this equality, one gets the orthogonality condition
$\left.\hat\chi_{_Y}\chi_{_R}\right|_{p_k=0} = \delta_{R,Y}$.

Formula (\ref{diff1}) follows from the definition of the skew Schur polynomials,
\be\label{Sw}
\chi_{_R}\{p+p'\}=\sum_Y\chi\{p'\}\chi_{_{R/Y}}\{p\}
\ee
and the Cauchy formula
\be
\exp\left(\sum_k \frac{z^kp_kp'_k}{k} \right) = \sum_Y z^{|Y|}\chi_{_Y}\{p\}\chi_{_Y}\{p'\}
\label{Cauchy}
\ee
Indeed,
\be\label{14}
\chi_{_R}\{p+p'\}=\exp\Big(\sum_kp_k'{\p\over\p p_k}\Big)\chi_{_R}\{p\}=\sum_Y\chi_{_Y}\{p'\}\hat\chi_{_Y}\{p\}\chi_{_R}\{p\}
\ee
and comparing the r.h.s. of (\ref{Sw}) and (\ref{14}), one obtains (\ref{diff1}).

Similarly, from the generalization of the Cauchy formula
\be
\exp\left(\sum_k \frac{z^kp_kp'_k}{k} \right)\chi_{_P}\{p'\} = \sum_Y z^{|Y|}\chi_{_{Y/P}}\{p\}\chi_{_Y}\{p'\}
\label{Cauchy2}
\ee
it follows that
\be\label{diff2}
\hat\chi_{_{Y/P}}\chi_{_R}\{p\} = \sum_{Q}N^{Y}_{PQ}\chi_{_{R/Q}}\{p\}
\ee
where $N^{Y}_{PQ}$ are the Littlewood-Richardson coefficients.

The r.h.s. of (\ref{bilH}) is a bilinear combination of the Schur and skew Schur polynomials,
and one needs to adjust $\xi_{_{X,Y}}$ so that the r.h.s. is again $\chi_{_R}\{p\}$.
It is not that easy, for instance, if one puts $\xi_{_{X,Y}}=\delta_{_{X,Y}}$,
then the r.h.s. is
$\sum_{X }  \chi_{_X}\{p\}\,\chi_{_{R/X}}\{p\} = \chi_{_R}\{2p\}\neq \chi_R\{p\}$.
Looking at actual $\xi_{_{X,Y}}$ for either the cut-and-join operators or the Ruijsenaars Hamiltonians,
one can observe a peculiar {\it hook} structure.
For the case of $\hat W$, see \cite{MMsumrules},
and now we show how this works \cite{triangle} for (\ref{RHam1}).
Applying the Cauchy formula (\ref{Cauchy}) (see \cite{Mcauchy} for a recent review)
to the two exponentials in (\ref{RHam1}), one gets
\be
\xi_{_{X,Y}}^{H_1} =
\chi_{_X}\{p_k = 1-t^{-2k}\}\cdot\chi_{_Y}\{p_k=q^{2k}-1\}\cdot\delta_{|X|,|Y|}
\label{xiRHam1}
\ee
Note that equal are only the {\it sizes} of the two Young diagrams $X$ and $Y$,
and that $\xi$ is {\it factorized} into the $X$- and $Y$-dependent pieces.
Additionally \cite{triangle}, at the peculiar locus $p_k = \{t^k\}$, the Schur polynomials are non-vanishing
only for the single-hook diagrams $X=[a+1,1^b]$:
\be
\chi_{_X}\Big\{p_k   =\{t^k\}\Big\}
=\sum_{Z\subset X} \chi_{_{X/Z}}\{p_k=t^k\}\cdot\chi_{_Z}\{p_k=-t^{-k}\}=
\sum_{Z\subset X,Z'} N_{_{Z,Z'}}^X\chi_{_{Z'}}\{p_k=t^k\}\cdot (-1)^{|Z|}\chi_{_{Z^\vee}}\{p_k=t^{-k}\}=\nn
\ee
\be
=\sum_{a,b} N_{_{[1^b],[a]}}^Xt^a\cdot (-1)^bt^{-b}
= \sum_{a,b}(-)^b\{t\}\cdot t^{a-b}\cdot\delta_{X,[a+1,1^b]}
\label{singhook}
\ee
where we used that $\chi_{_{X/Z}} = \sum_{Z'}N_{_{Z,Z'}}^X\chi_{_{Z'}}$ and $[a]\otimes[1^b]=[a,1^b]\oplus
[a+1,1^{b-1}]$ for the decomposition of the tensor product of representations of the linear group. Here $Z^\vee$ denotes the transposition of the Young diagram $Z$.

Substituting (\ref{singhook})  into  (\ref{xiRHam1}), one obtains
\be
\xi_{_{X,Y}}^{H_1} =\sum_{a,b,c,d}\{q\}\{t\}\cdot  (-)^{b+d}\cdot \frac{q^{2c+1}}{t^{2b+1}}\cdot\delta_{a+b-c-d}
\cdot\delta_{X,[a+1,1^b]}\cdot\delta_{Y,[c+1,
1^d]}
\label{xiRHam2}
\ee
i.e.
\be
\frac{t}{q}\cdot \frac{\hat H_1-1}{\{q\}\{t\}}\,\chi_{_R}
= \frac{t^2}{q^2-1}\cdot \frac{\hat H_1-1}{t^2-1}\,\chi_{_R}
\ = \ \chi_{_{[1]}}\chi_{_{R/[1]}}
+ \frac{q}{t} \Big(t\,\chi_{_{[2]}}-\frac{\chi_{_{[1,1]}}}{t}\Big)
\Big(q\,\chi_{_{R/[2]}}-\frac{\chi_{_{R/[1,1]}}}{q}\Big)
+ \nn \\
+\Big(\frac{q}{t}\Big)^2\Big(t^2\chi_{_{[3]}}-\chi_{_{[2,1]}}
+\frac{\chi_{_{[1,1,1]}}}{t^2}\Big)
\Big(q^2\chi_{_{R/[3]}}-\chi_{_{R/[2,1]}}+\frac{\chi_{_{R/[1,1,1]}}}{q^2}\Big)
+ \nn \\
+\Big(\frac{q}{t}\Big)^3\Big(t^3\chi_{_{[4]}}-t\,\chi_{_{[3,1]}}
+\frac{\chi_{_{[2,1,1]}}}{t}
-\frac{\chi_{_{[1,1,1,1]}}}{t^3}\Big)
\Big(q^3\chi_{_{R/[4]}}-q\,\chi_{_{R/[3,1]}}+\frac{\chi_{_{R/[2,1,1]}}}{q}-
\frac{\chi_{_{R/[1,1,1,1]}}}{q^3}\Big)
+ \ldots
\label{RHam1hook}
\ee
Note that $\chi_{_{[2,2]}}$ and $\chi_{_{R/[2,2]}}$ are absent in the last line.

For $q=t$, there are interesting sum rules saying that the r.h.s. is proportional to $\chi_{_R}$.
For $q\neq t$, this remains true only for antisymmetric $\chi_{[1^s]}$:
in this case, only the last terms in each second bracket contribute,
and the $q$-dependence is immediately eliminated.
A non-trivial sum rule is, however, still needed:
$$
\chi_{_{[1]}}\chi_{_{[1]^{s-1}}}\!
-  \Big( \chi_{_{[2]}}-\frac{\chi_{_{[1,1]}}}{t^2}\Big)\chi_{_{[1]^{s-2}}}\!
+ \Big( \chi_{_{[3]}}-\frac{\chi_{_{[2,1]}}}{t^2}
+\frac{\chi_{_{[1,1,1]}}}{t^4}\Big)\chi_{_{[1]^{s-3}}}\!
- \Big( \chi_{_{[4]}}-\frac{\chi_{_{[3,1]}}}{t^2}
+\frac{\chi_{_{[2,1,1]}}}{t^4}
-\frac{\chi_{_{[1,1,1,1]}}}{t^6}\Big)\chi_{_{[1]^{s-4}}}\!
+ \ldots =
$$
\vspace{-0.5cm}
\be
=  ( 1+t^{-2}+\ldots +t^{2-2s})\cdot \chi_{_{[1]^s}}
= \frac{1-t^{-2s}}{1-t^{-2}}\cdot \chi_{_{[1]^s}}
\ee
Note that the empty diagram contributes somewhat differently,
because it is associated with the zero hook, not unit, and
the corresponding contribution is removed from the l.h.s. of (\ref{RHam1hook}).

\bigskip

{\bf 5.} In fact,  the
factorization of $\xi_{_{X,Y}}$ in (\ref{bilH}),
\be
\xi^{\rm exp}_{_{X,Y}}=\xi^L_{_X}\cdot\xi^R_{_Y}
\ee
is a general feature of exponential Hamiltonian.
It takes place at any choice of ``background" times $\alpha_k$ and $\beta_k$
not obligatory equal to $1-t^{-2k}$ and $q^{2k}-1$
(what happens at these particular values is an additional restriction to
the {\it single}-hook diagrams $X$ and $Y$).
One can increase the rank of $\xi_{_{X,Y}}$ to an arbitrary value $M$
by taking a sum of exponential Hamiltonians:
$$
\hat H  = \sum_{i=1}^M \oint\frac{dz}{z}
 \exp\left(\sum_{k>0}{\alpha^{(i)}_k\,p_kz^k\over k}\right)
 \exp\left(\sum_{k>0} {\beta^{(i)}_k\over z^k}{\partial\over\partial p_k}\right)
\ \ \  \Longleftrightarrow    \ \ \
$$
\vspace{-0.4cm}
\be
\ \ \  \Longleftrightarrow    \ \ \
\xi^{\hat H}_{_{X,Y}}  = \delta_{|X|,|Y|} \cdot
\sum_{i=1}^M \chi_{_X}\{p_k=\alpha^{(i)}_k\} \cdot \chi_{_Y}\{p_k=\beta^{(i)}_k\}
\label{rankM}
\ee

Another general feature of the exponential Hamiltonians is that the
``chiral components" of $\xi_{_{X,Y}}$ are the Schur polynomials and therefore
they are constrained by relations
\be
\chi_{_X}\{\alpha\}\chi_{_{X'}}\{\alpha\}
= \sum_{Z\in X\otimes X'}N^Z_{XX'}\chi_{_Z}\{\alpha\}
\label{prodchar}
\ee
More sophisticated constraints
of the same origins are imposed
on $\xi_{_{X,Y}}$
if $\hat H$ is a {\it multi}-linear combination of exponentials
like a higher Ruijsenaars Hamiltonian.

\bigskip

{\bf 6.} The higher Ruijsenaars Hamiltonians $\hat H_m$
are acting as bilinears with only $m$-hook diagrams contributing.
Indeed, as explained in \cite{MMgenmac}, these Hamiltonians are made from
the polylinear combinations of different harmonics of the exponential operator
\be\label{Vm}
\hat V_m(z) =
\exp\left(\sum_{k>0}{(1-t^{-2mk})\,p_kz^k\over k}\right)\cdot
 \exp\left(\sum_{k>0}
 \frac{t^{2mk}-1}{t^{2k}-1}{q^{2k}-1\over z^k}
{\partial\over\partial p_k}\right)=\nn\\
=\sum_{X,Y}  z^{|X|-|Y|} t^{(m-1)|Y|-|X|}q^{|Y|} \cdot
\chi_{_X}\Big\{p_k=\{t^k\}\Big\}
\chi_{_Y}\!\left\{p_k=\frac{\{t^{mk}\}\{q^k\}}{\{t^k\}}\right\}
\cdot \chi_{_X} \hat\chi_{_Y}
\ee
Now $\chi_{_X}\!\left\{p_k=\frac{\{t^{mk}\}\{q^k\}}{\{t^k\}}\right\} = 0$ for the diagrams $X$ with more than $m$ hooks and we need a generalization of (\ref{singhook}). It can be obtained using formulas of the Appendix.
Then, the higher Hamiltonians $\hat H_m$ are fully localized at diagrams with no more than $m$-hooks,
\be
\boxed{
 \hat H_m = \!\!\!\!\! \sum_{\stackrel{X,Y:}
{{\rm hook}_X, {\rm hook}_Y \leq  m}} \!\!\!\!\!  \delta_{|X|,|Y|}\cdot \xi_{_{X,Y}}
\cdot \chi_{_X} \cdot \hat\chi_{_Y}
}
\label{Hm}
\ee

For instance, while the main Hamiltonian (\ref{RHam1}) is just $\hat H_1=\oint\frac{dz}{z}\hat V_1(z)$,
the second one is
\be\label{H2}
\hat H_2 = \oint\frac{dz}{z}\hat V_2(z) - \frac{\{t^2\}}{\{t\}^2}
\oint\frac{dz_1}{z_1}\oint\frac{dz_2}{z_2}\frac{(z_1-z_2)^2}{(z_1-t^2z_2)(z_1-t^{-2}z_2)}
\,:\hat V_1(z_1)\hat V_1(z_2):
\ee
If we expand it into characters as before, at most {\it two}-hook diagrams contribute
in the both terms. Indeed, this is the case for the first term because of (\ref{Vm}). As for the second term, we
note that
\be\label{VV}
:\hat V_1(z_1)\hat V_1(z_2):=\sum_{X_1,Y_1,X_2,Y_2}  z_1^{|X_1|-|Y_1|}z_2^{|X_2|-|Y_2|} q^{|Y_1|+|Y_2|}t^{-|X_1|-|X_2|} \times\nn\\
\times\chi_{_{X_1}}\Big\{p_k=\{t^k\}\Big\}\chi_{_{X_2}}\Big\{p_k=\{t^k\}\Big\}
\chi_{_{Y_1}}\!\left\{p_k=\{q^k\}\right\}\chi_{_{Y_2}}\!\left\{p_k=\{q^k\}\right\}
\cdot \chi_{_{X_1}}\chi_{_{X_2}} \hat\chi_{_{Y_1}}\hat\chi_{_{Y_2}}=\nn\\
=\sum_{a_i,b_i,c_i,d_i}  (-1)^{d_1+d_2+1}
z_1^{a_1+b_1-c_1-d_1}z_2^{a_2+b_2-c_2-d_2} q^{2c_1+2c_2+2}(-t^{-2})^{b_1+b_2+1}\{q\}^2\{t\}^2\times\nn\\
\times N_{_{[a_1+1,1^{b_1}],[a_2+1,1^{b_2}]}}^X N_{_{[c_1+1,1^{d_1}],[c_2+1,1^{d_2}]}}^Y\chi_{_{X}} \hat\chi_{_{Y}}
\ee
Since the tensor product of two representations associated with 1-hook Young diagrams contains not more than 2-hook Young diagrams, $X$ and $Y$ in this formula are also not more than 2-hook. However, though in each of the two terms $\xi_{_{X,Y}}$ in (\ref{bilH}) is factorized to the product $\xi^L_{_X}\cdot\xi^R_{_Y}$, in the sum, it is not. One can use formulas (\ref{h1}) and (\ref{h2}) from the Appendix in order to evaluate the first term in (\ref{H2}), and (\ref{VV}) and integral (\ref{int}) in the Appendix in order to evaluate the second term and obtain $\xi_{_{X,Y}}$.

\bigskip

{\bf 7.} After these examples, we can return to (\ref{bilH}) and ask
how to find $\xi_{_{X,Y}}$ if the Hamiltonian is a priori unknown.
We begin from the question, what is the Hamiltonian if the eigenfunctions
$\Psi_I$ are already known.
The formal answer is
\be
\hat{\cal H} = \sum_I E_I\cdot \Psi_I \hat \Psi_I
\ \ \ \Longleftrightarrow \ \ \
\hat{\cal H}\,\Psi_I = E_I\cdot \Psi_I
\label{obviousHam}
\ee
where $\hat \Psi_I$ is a dual {\it operator} with the property
$\hat \Psi_I \Psi_J = \delta_{IJ}$.
Our goal in sections {\bf 7}--{\bf 10}
is to make this formula a little more explicit for the case of symmetric functions.

This question makes sense already at the Schur level, but we consider it directly for the
Kerov functions, because there is no much difference:
in any case, we need to return to the Schur case in sec.{\bf 10}.
We remind from \cite{MMkerov} that
\be
\Ker_R\{p\} =\sum_{R'} {\cal K}^{(g)}_{_{R,R'}}\cdot \chi_{_{R'}}\{p\}
\label{Kerexpan}
\ee
The sum is actually triangular and goes over $R'\leq R$ w.r.t. lexicographic
or inverse lexicographic ordering.
The two orderings are not equivalent beyond the Macdonald locus (\ref{Macloc})
and define two {\it dual} sets of
Kerov functions, $\Ker\{p\}$ and $\widetilde\Ker\{p\}$
(they actually deviate from each other starting from level $6$,
where, for example, $\gamma$ in (\ref{KLRvio}) vanishes for $\Ker$,
but not for $\widetilde\Ker$).
Concrete entries of the triangular Kostka-Kerov matrices
${\cal K}^{(g)}$ and $\widetilde{\cal K}^{(g)}$
can be easily calculated by orthogonalization method w.r.t. the scalar product (\ref{pprod}),
and they have interesting properties as functions of Kerov times $\{g_k\}$.
We assume them known, see \cite{MMkerov2} for some examples.
Then what we need are the relations
\be
\hat{\cal H}\,\Ker_R = E_R\cdot \Ker_R
\ \ \ \Longleftrightarrow \ \ \
\sum_{R',X,Y} {\cal K}_{_{R,R'}}\cdot
\left( \xi_{X,Y}\cdot \chi_{_X}\chi_{_{R'/Y}} - E_R\cdot\chi_{R'}\right) = 0
\label{condforxi}
\ee
If we assume that the sizes of $X$ and $Y$ remain the same,
as it was in the case of (\ref{RHam1}), $\xi_{_{X,Y}}\sim \delta_{|X|,|Y|}$,
then at each level $n$, where we have $\sigma_n$ Young diagrams of the size $n$
($\sigma$'s can be obtained from $\sum_n \sigma_n q^n = \prod_n (1-q^n)^{-1}$),
there are $\sigma_n^2$ coefficients $\xi_{_{X,Y}}$ and exactly the same number
of equations from (\ref{condforxi}).
Indeed, there are $\sigma_n$ choices for $R$, and the equation is a polynomial of $p$,
i.e. vanishing should be coefficients in front of all the $\sigma_n$ monomials $p_\Delta$.
Actually, counting is a little less direct, because (\ref{condforxi}) contains
contributions from $\xi_{_{X,Y}}$ with $|X|,|Y|\leq |R|$, but $\xi_{_{X,Y}}$
with smaller $X$ and $Y$ are defined in consideration of smaller $R$.
In result, we have equal numbers of variables and equations, and this means that
$\xi_{X,Y}$ can be unambiguously deduced from (\ref{condforxi}).
This can be done {\it for any} given set of eigenvalues $\{E_R\}$,
i.e. we have an $\sum_n \sigma_n$-parametric set of Hamiltonians $\hat{\cal H}_Q$
labeled essentially by Young diagrams rather than just by an integer:
\be
\boxed{
\hat{\cal H}_Q\, \Ker_R =  \Ker_R \cdot \delta_{R,Q}
}
\label{HamQ}
\ee
The multiplicity nicely matches that of $\hat W$ operators.
In terms of these operators, the Hamiltonian (\ref{condforxi})
with a given set of eigenvalues is
\be
\hat{\cal H} = \sum_Q E_Q\hat{\cal H}_Q
\label{HvsHQ}
\ee
An explicit example of this construction for the first three levels can be found in the Appendix.

\bigskip

{\bf 8.}
Hamiltonians $\hat {\cal H}_Q$ explicitly respect triangularity of the expansion
(\ref{Kerexpan}).
To see this, introduce the set of $g$-independent operators $\hat h_Q$
with the property
\be
\hat h_{_Q} \chi_{_R}  = \delta_{R,Q}
\label{defh}
\ee
which are actually counterparts of $\hat{\cal H}_Q$ for the Schur polynomials,
$\hat h_Q =   \chi_{_Q}^{-1}\cdot\left.\hat {\cal H}_Q\right|_{g_k=1}$.
Now, the triangularity implies, for instance, that
$\chi_{_[r]}$ in symmetric representation $[r]$
appears only in the highest Kerov function $\Ker_{[r]}$,  and does not contribute
to the expansion of all other $\Ker_Q$ at the same level $|Q|=r$.
Therefore
\be
\hat{\cal H}_{[r]} =   \Ker_{[r]}\cdot \hat h_{[r]}
\ee
since it is sufficient for the operator $\hat h_{[r]}$ to annihilate all {\it Schur} polynomials,
except for $\chi_{_{[r]}}$, thus it can be (and is) independent of the $g$-variables.
However, the next operator $\hat{\cal H}_{[r-1,1]}$  should annihilate not just $\chi_{_{[r]}}$
but a $g$-dependent combination of $\chi_{_{[r]}}$ and $\chi_{_{[r-1,1]}}$,
which enters $\Ker_{[r]}$, and thus it needs to depend on $g$.
However, the only Kerov function, which contains $\chi_{_{[r-1,1]}}$,
and differs from $\Ker_{[r-1,1]}$ is
$\Ker_{[r]} = \chi_{_{[r]}} + {\cal K}_{[r],[r-1]}^{(g)}\chi_{_{[r-1,1]}} + \ldots$,
thus
\be
\hat{\cal H}_{[r-1,1]} =   \Ker_{[r-1,1]} \cdot
\Big( \hat h_{[r-1,1]} -{\cal K}_{[r],[r-1]}^{(g)} \cdot\hat h_{[r]}\Big)
\ee
An explicit example of this phenomenon is the coincidence of two underlined operators
in (\ref{H2andH11}), as well as the coefficient in front of the last term in
the second expression for $\hat{\cal H}_{[2,1]}$.
In general, ${\cal H}_Q$ are related to $\hat h_Q$ by an {\it upper} triangular
transformation with the transposed inverse of the Kostka-Kerov matrix:
\be
\boxed{
\hat{\cal H}_Q =  \Ker_Q\cdot\sum_{S\geq Q} {\cal K}^{-1}_{SQ}\, \hat h_S
}
\label{HamQvsh}
\ee
Indeed, then
\be
\hat{\cal H}_Q \Ker_R
=   \Ker_Q\cdot \sum_{S,T} {\cal K}^{-1}_{SQ} {\cal K}_{RT}\hat h_S \chi_{_T}
=  \Ker_Q \sum_{S} {\cal K}^{-1}_{SQ}{\cal K}_{RS}
= \Ker_R\cdot \delta_{R,Q}
\ee
One can substitute expansion (\ref{Kerexpan}) into (\ref{HamQvsh}),
which provides
\be
\hat{\cal H}_Q =  \sum_{S,T}
{\cal K}^{-1}_{SQ}{\cal K}^{\phantom{-1}}_{QT}\cdot \chi_{_T}\hat h_{_S}
\label{HQvsKK}
\ee
There is no sum over $Q$.
Performing the sum, one gets the identity operator
\be
\sum_Q  \hat{\cal H}_Q  = \sum_S \chi_{_S}\hat h_{_S}
\label{simplestsumrule}
\ee
which leaves every Schur polynomial, and hence every Macdonald and Kerov ones, intact:
\be
\sum_S \chi_{_S}\hat h_{_S}\, \Ker{_R} = \Ker{_R}
\ee
For the dual Kerov functions, one gets their own Hamiltonians
\be
\hat{\!{\widetilde {\cal H}}}_Q =
 {\widetilde\Ker}_Q\cdot\sum_S { {{\cal K}}}^{-1}_{SQ}\, \hat h_S
\ee
and there are obvious operators which convert $\Ker$ into $\widetilde\Ker$ and back:
\be
\hat{ {\cal T}}_Q =  \widetilde \Ker_Q\cdot\sum_S  {{\cal K}}^{-1}_{SQ}\, \hat h_S
\ \ \Longleftrightarrow \ \
\hat{ {\cal T}}_Q \,\Ker_R = \delta_{R,Q} \cdot\widetilde\Ker_R
\nn \\
\hat{\!{\widetilde{\cal T}}}_Q =   \Ker_Q\cdot\sum_S \tilde{{\cal K}}^{-1}_{SQ}\, \hat h_S
\ \ \Longleftrightarrow \ \
\hat{\!{\widetilde{\cal T}}}_Q\,\widetilde\Ker_R = \delta_{R,Q} \cdot \Ker_R
\ee

\bigskip

{\bf 9.} Our next task is to construct the $g$-independent operators $\hat h_S$ explicitly.
Note that $\hat h_{_S} = \hat\chi_{_S} + \ldots$
with non-trivial corrections, because we do not put all $p_k=0$ in (\ref{defh}),
thus the standard orthogonality condition, mentioned in the first paragraph of sec.{\bf 4}
is not enough.

Already the very first operator $\hat h_{[1]}$ has quite an inspiring form
clearly seen in the first line of (\ref{H2andH11}):
\be
\hat h_{[1]} =   \sum_X (-)^{|X|}\chi_{_X}\cdot
\left(\sum_{Y\in X^\vee\otimes[1]} \hat\chi_{_{Y}}\right)
\ee
Generalization is obvious, and it is indeed true:
for arbitrary $Q$
\be
\boxed{
\hat h_{Q} =   \sum_X (-)^{|X|}\chi_{_X}\cdot
\left(\sum_{Y\in X^\vee\otimes Q} \hat\chi_{_{Y}}\right)
}
\label{hQ}
\ee
To prove (\ref{defh}), one can apply the Cauchy formula (\ref{Cauchy}) in the form
\be
\exp\left(-\sum_k {p_k }\frac{\p}{\p p'_k}\right)
= \sum_X (-)^{|X|}\chi_{_X}\{p\}\hat\chi_{_{X^\vee}}\{p'\}
\ee
to
\be
\chi_{_R}\{p+p'+p''\} = \sum_Y \chi_{_{R/Y}}\{p\}\chi_{_{Y}}\{p'+p''\}
= \sum_{Y,Q}  \chi_{_{R/Y}}\{p\}\chi_{_{Y/Q}}\{p'\}\chi_{_Q}\{p''\}
\ee
Then at the l.h.s., we get a shift of $p'$ by $-p$, and putting $p'=0$ afterwards
reduces it to $\chi_{_R}\{p''\}$.
Comparison with the r.h.s. gives:
\be
\chi_{_R}\{p''\} =
\sum_{X,Y,Q} (-)^{|X|}\chi_{_X}\{p\}\chi_{_{R/Y}}\{p\}
\chi_{_Q}\{p''\} \left.\Big(\hat\chi_{_{X^\vee}}\{p'\} \chi_{_{Y/Q}}\{p'\}\Big)\right|_{p'=0}
\ee
The last bracket imposes the condition that $Y\in Q\otimes X^\vee$,
and, comparing the terms with $\chi_{_Q}\{p''\}$ at both sides, we obtain the desired relation
\be
\hat h_{_Q}\, \chi_{_R}\{p\} =
\sum_X (-)^{|X|}\chi_{_X}\{p\}\sum_{Y\in X^\vee\otimes Q} \chi_{_{R/Y}}\{p\}
= \delta_{Q,R}
\ee
Note that the contributing to the sum at $Q=R$ is just the term with $X=\emptyset$.

{\bf Eqs.(\ref{HamQvsh}) and (\ref{hQ}) give a complete explicit construction
of Hamiltonians for the Kerov functions}
(and in fact for any system of symmetric functions
defined by a linear transformation of Schur polynomials with the matrix ${\cal K}$).

This is, however, not yet the case when the dream came true.
The naive Hamiltonians (\ref{HamQvsh}) depend explicitly on the Kostka-Kerov matrix,
and can not be used to {\it derive} it.
At the same time, in the Macdonald case, there were {\it very special}
Ruijsenaars Hamiltonians (\ref{RHam1}), which do not refer to the Kostka matrix,
and could be used for its derivation.
Despite this is technically much harder than using the orthogonalization procedure,
still it is conceptually important that {\it such} Hamiltonians exist.
We do not discuss here what is so special about (\ref{RHam1}) and what
are the chances to find {\it their} counterparts in the Kerov case.

\bigskip

{\bf 10.}
The Ruijsenaars Hamiltonian (\ref{RHam1}) was described by a maximally
degenerate (factorized) matrix $\xi_{_{X,Y}}$, but instead it had no free parameters
in the set of eigenvalues, i.e. even in Macdonald case it was some peculiar
combination of our Hamiltonians $\hat{\cal H}_Q$.
In fact, to get the exponential Hamiltonian (\ref{RHam1}),
one should just substitute the eigenvalues (\ref{evRMac})
into (\ref{HvsHQ}) and (\ref{HQvsKK}) and restrict the Kostka-Kerov matrix ${\cal K}$
to the Macdonald locus.
However, since eigenvalues depend on $Q$, one needs
a generalization of the sum rule (\ref{simplestsumrule}).
Our next goal is to reveal in the simplest example of the Appendix what is a peculiar
combination of  $\hat{\cal H}_Q$ leading to (\ref{RHam1}), and to explain why the same
factorizability (rank one) condition can not be imposed outside the Macdonald locus
(\ref{Macloc}).
We can also look at the weakened, say, rank-two condition and find what is the
corresponding extension of Macdonald locus.
In fact, matrix $\xi_{_{X,Y}}$ is not of rank 1 (not fully factorized)
already at level 2, see (\ref{xilev2}).
However, one can make it degenerate by adjusting one of the three eigenvalues:
\be
E_{[1,1]} = \frac{2g_1^2E_{[2]}E_{[1]}}{(g_{2}+g_{1}^2)E_{[2]}-2g_2E_{[1]}}
\label{lev2fact}
\ee
One can repeat this trick at level $3$, then all the three eigenvalues get
expressed through $E_{[1]}$ and $E_{[2]}$:
\be
E_{[1,1,1]} = \frac{3g_1(g_1^2+g_2)E_{[1]}}
{ (2g_3+3g_2g_1+g_1^3)E_{[2]} + (g_1^3-3g_2g_1-4g_3)E_{[1]}}\cdot E_{[2,1]}
\nn \\
E_{[2,1]} = \frac{( g_1^2+g_2)E_{[2]}^2 + (g_1^2-3g_2)E_{[2]}E_{[1]}+2g_2E_1^2}
{( g_1^2+g_2)E_{[2]}-2g_2E_1}
\nn \\
E_{[3]} =
\frac{3g_1(g_1^2+g_2)\Big(( g_1^2+g_2)E_{[2]}^2 + (g_1^2-3g_2)E_{[2]}E_{[1]}+2g_2E_1^2\Big)}
{(g_3g_2^2-4g_2^2g_1^3+3g_3g_1^4)E_{[2]}+(6g_3g_2g_1^2-2g_3g_2^2+8g_2^2g_1^3)E_{[1]}}
\label{lev3fact}
\ee
Moreover, at this locus (in $E$-space) we get quite a nice factorized formula
\be
\xi^{\rm fact}_{_{[a+1,1^b],[c+1,1^d]}} \ \stackrel{?}{=}\
(-)^{c+d} \cdot\frac{E_{[1^{b+1}]}-E_{[1^{b}]}}{E_{[1]}}\cdot
(E_{[c+1]}-E_{[c]})
\label{xifact}
\ee
which reproduces (\ref{RHam1}) at Macdonald locus (\ref{Macloc}) in the $g$-space,
but does {\it not} work beyond it, starting from level $4$,
where a two-hook diagram emerges for the first time.
Let us emphasize again that (\ref{xifact}) should be considered
not {\it freely}, but for the eigenvalues
restricted by conditions (\ref{lev2fact}), (\ref{lev3fact}), etc,
which, on the Macdonald locus, reduce to  (\ref{evRMac}).
{\bf Beyond the Macdonald locus, eq.(\ref{xifact}) has no clear value},
at least the Kerov functions are {\it not}
the eigenfunctions of the operators that one can build from this $\xi^{\rm fact}$.
Note also that even if (\ref{xifact}) would be true, it satisfies (\ref{rankM})
with $M=1$ but not obligatory (\ref{prodchar}),
i.e. the necessary conditions for an exponential Hamiltonian to exist would not
be fulfilled.

\bigskip

{\bf 11.}
The question, however, remains, if one can get an exponential Hamiltonian with $M>1$.
One option here is to look for generalizations of the Macdonald locus (\ref{Macloc})
with the hope that some factorization properties survive.
Since  the vanishing property (\ref{singhook}) played a role in construction of
exponential Hamiltonians, it is instructive that it has a generalization to other loci:
\be
\chi_{_R}\left\{p_k =\frac{\{t^{mk}\}}{\{t^k\}}\cdot\{q^k\}\right\}= 0
&{\rm for} & {\rm hook}_R>m \nn \\
\chi_{_R}\left\{p_k = \prod_{a=1}^m\{q_a^k\} \right\} = 0
&{\rm for} & {\rm hook}_R>2^{m-1}
\ee
which is a simple corollary of the general theorem: the Schur polynomial $\chi_{_R}$ is non-zero at $p_k=\sum_{i=1}^Nx_i^k-\sum_{i=1}^Ny_i^k$ iff $R$ has no more than $N$ hooks. One of the simplest ways to prove this theorem is to realize such a Schur polynomial as a fermionic average of a product of fermions: $\chi_{_R}=\Big<\prod_{i=1}^N\psi^*(y_i)\psi(x_i)\cdot
\prod_{a}^N\psi_{-\mu_a}^*\psi_{\nu_a}\Big>$, where $\mu_a$ are lengths of the vertical hook legs, and $\nu_a+1$ are lengths of the horizontal ones \cite{KM}. One can also prove the theorem using the hook determinant formula from the Appendix.

As already mentioned in sec.{\bf 6},
the first series $p_k =\frac{\{t^{mk}\}}{\{t^k\}}\cdot\{q^k\}$
appears in study of the higher Ruijsenaars Hamiltonians.
The second series $p_k = \prod_{a=1}^m\{q_a^k\}$
is naturally relevant for considerations at the Freund-Zabrodin \cite{FZ} locus
\be
g_k^{FZ} = \frac{\{q_a^k\}}{\{t_a^k\}}
\label{FZloc}
\ee
in the space of Kerov times $g_k$.
We leave a detailed analysis of this possibility for the future.

\bigskip

{\bf 12.} Another question which we mentioned in the beginning of this text is
if one can define generalized functions with the help of $\hat{\cal H}_Q$?
Of  course, for any system of, say, two-point generalized functions
\be
\Ker_{Q,Q'}\{p,p'\} = \sum_{(S,S')\leq (Q,Q')} {\cal K}_{Q,Q'|S,S'}\cdot
\chi_{_S}\{p\}\chi_{_{S'}}\{p'\}
\ee
defined by a triangular transformation of the bi-linear Schur basis,
there is a direct generalization of (\ref{HvsHQ}) and (\ref{HamQvsh}):
\be
\hat{\cal H} = \sum_{Q,Q'} E_{Q,Q'}\cdot \hat{\cal H}_{Q,Q'}
\ee
\be
\hat{\cal H}_{Q,Q'} = \Ker_{Q,Q'}
\sum_{(S,S')\geq (Q,Q')} {\cal K}_{S,S'|Q,Q'}^{-1} \hat h_S\{p\}\hat h_{S'}\{p'\}
\ee
which defines a Hamiltonian for arbitrary set of the eigenvalues:
\be
\hat{\cal H} \,\Ker_{Q,Q'} =   E_{Q,Q'}\cdot \Ker_{Q,Q'}
\ee
The question is, however, to find a restricted sub-set of Hamiltonians
which could be described with no explicit reference
to the generalized Kostka-Kerov matrix ${\cal K}$
and thus could be used to {\it define} it.

In the Macdonald case, such {\it interesting} Hamiltonians exist \cite{MMgenmac},
and are given by simple sums of Ruijsenaars exponential Hamiltonians
with a simple triangular mixing of time sets $\{p\}$ and $\{p'\}$.
More precisely, the first one is obtained from (\ref{RHam1}) and (\ref{xiRHam1})
by the transformation
\be
\hat{H}_1\{p\}=\sum_{X,Y} \xi^{H_1}_{_{X,Y}}\chi_{_X}\{p\}\hat\chi_{_Y}
\ \longrightarrow \
\hat{ H_1}\{p,p'|Q\} = \sum_{X,Y} \xi^{H_1}_{_{X,Y}}\left(\chi_{_X}\{p\}\hat\chi_{_Y}\{p\}
 + \frac{1}{A^2}\cdot\chi_{_X}\{p'+\epsilon p\}\hat\chi_{_Y}\{p'\}\right)
\ee
with the deformation parameter $A^{-2}$.
At the Macdonald locus, the mixing is
$p'_k+\epsilon_kp_k$ with
$\epsilon_k =  1 - \Big(\frac{t}{q}\Big)^{2k} $  made from the third item of the
DIM triple $q,t^{-1},tq^{-1}$ \cite{DIMMac}.

\bigskip

{\bf 13.}
To conclude, in this paper we constructed {\it a full set} of
Hamiltonians for Kerov functions.
This is a superficially large set, and it can not help to define the
functions {\it per se}, because our Hamiltonians explicitly contain the
Kostka-Kerov matrix, i.e. they use it as an input rather than serve as
a tool to {\it define} the Kerov functions.
In other terms, they demonstrate super-integrability, but
lack the advantage of (\ref{RHam1}) and its relatives,
which formed a smaller set of operators depending on parameters $q$ and $t$
in a simple explicit way, not through the Kostka-Macdonald matrix,
and thus could serve its definition.
The fact that the peculiar properties of the Hamiltonians in the Schur case have generalizations to
other loci gives a hope to lift the construction, say, to the
Freund-Zabrodin generalizations of the Macdonald locus, but this is beyond the scope
of the present paper.
Our main goal was to demonstrate that the notion of Kerov Hamiltonians
has a clear meaning,
and to make a setting for the next attacks on this interesting problem.

\section*{Acknowledgements}

\noindent
This work was supported by the Russian Science Foundation (Grant No.16-12-10344).

\section*{Appendix}

\paragraph{Schur polynomials as determinants of the hook constituents.}

Here we write down the determinant representation of the Schur polynomials in terms of hook constituents following \cite{KM,DJKMIII}. These formulas are convenient for dealing with diagrams with a restricted number of hooks.

First of all, let us note that, as follows from the Cauchy formula (\ref{Cauchy}),
\be
\exp\Big(\sum_k{p_kz^k\over k}\Big)=\sum_n\chi_{[n]}z^n
\ee
On the other hand, since $N_{[a],[1^]}^X=\delta_{X,[a,1^b]}+\delta_{X,[a+1,1^{b-1}]}$ and $\chi_{_X}\{p_k\}=
(-1)^{|X|}\chi_{_{X^\vee}}\{-p_k\}$,
\be\label{1hook}
\chi_{[a,1^b]}=(-1)^{b+1}\sum_{j\ge 0}\chi_{[j+b+1]}\{-p_k\}\chi_{[a-j-1]}\{p_k\}
\ee
Then, for the Young diagram $R$ consisting of $n$ hooks with vertical leg length $b_i+1$, and horizontal leg lengths $a_i$, $i=1\ldots n$, the Schur polynomial reads
\be\label{deth}
\chi_{_R}=\det_{1\le i,j\le n}
\chi_{[a_i,1^{b_j}]}
\ee

\paragraph{An example: the second Macdonald Hamiltonian.}

Here we demonstrate in detail how the formulas work in the case of the second Macdonald Hamiltonian (\ref{H2}).

In the first term of (\ref{H2}), non-vanishing are contributions of 1-hook diagrams with\footnote{It can be obtained using (\ref{1hook}) from
\be
\chi_{[a]}\Big\{p_k=(t^k+t^{-k})\{q^k\} \Big\}=q^a\ {\{t^{a+1}\}-{t+t^{-1}\over q^2}\{t^a\}+q^{-4}\{t^{a-1}\}\over\{t\}}
\ee
In the simplest way, this latter is obtained from the generating function of symmetric Young diagrams in this case:
\be
\sum_az^a\cdot\chi_{[a]}\Big\{p_k=\sum_i(x_i^k-y_i^k) \Big\}=\prod_i{1-y_iz\over 1-x_iz}
\ee
i.e.
\be
\chi_{[a]}\Big\{p_k=\sum_i(x_i^k-y_i^k) \Big\}={x_1^a(x_1-(y_1+y_2)+y_1y_2/x_1)-x_2^a(x_2-(y_1+y_2)+y_1y_2/x_2)\over
x_1-x_2}
\ee
with $x_1=qt$, $x_2=q/t$, $y_1=t/q$, $y_2=1/(qt)$.
}
{\footnotesize
\be\label{h1}
\chi_{_{[a_1+1, 1^{b_1}]}}\!\Big\{p_k=(t^k+t^{-k})\{q^k\} \Big\} =
 (-)^{b_1+1} q^{a_1 -b_1 }
\frac{\{qt\}\{q/t\}\Big(t^{a_1+b_1+1}+t^{-a_1-b_1-1}\Big)
- \{q\}\Big(\{qt\}t^{a_1-b_1}+\{q/t\}t^{-a_1+b_1}\Big) }{\{t\}^2}
\nn
\ee
}
and of the 2-hook diagrams with (using (\ref{deth}))
\be\label{h2}
\chi_{_{[a_1+1,a_2+1,2^{b_2},1^{b_1-b_2-1}]}}\!\Big\{p_k=(t^k+t^{-k})\{q^k\} \Big\} =
 (-)^{b_1+b_2} q^{a_1+a_2-b_1-b_2-3}\{q\}^2\{qt\}\{q/t\}
\frac{\{t^{a_1-a_2}\}\{t^{b_1+1-b_2}\}}{\{t\}^2}
\ee

As for the second term, the product of two $V_1$ is
{\footnotesize
\be
:\hat V_1(z_1)\hat V_1(z_2): \
= \left(1+ \{t\}\left(\frac{z_1}{t}\chi_{_{[1]}}
+ \frac{z_1^2}{t^2}\left(t\,\chi_{_{[2]}}-\frac{1}{t}\,\chi_{_{[1,1]}}\right)
+ \ldots\right)\right)
\left(1+ \{t\}\left(\frac{z_2}{t}\chi_{_{[1]}}
+ \frac{z_2^2}{t^2}\left(t\,\chi_{_{[2]}}-\frac{1}{t}\,\chi_{_{[1,1]}}\right)
+ \ldots\right)\right)\cdot
\nn \\
\cdot\left(1+ \{q\}\left(\frac{q}{z_1}\hat\chi_{_{[1]}}
+ \frac{q^2}{z_1^2}\left(q\,\hat\chi_{_{[2]}}-\frac{1}{q}\,\hat\chi_{_{[1,1]}}\right)
+ \ldots\right)\right)
\left(1+ \{q\}\left(\frac{q}{z_2}\hat\chi_{_{[1]}}
+ \frac{q^2}{z_2^2}\left(q\,\hat\chi_{_{[2]}}-\frac{1}{q}\,\hat\chi_{_{[1,1]}}\right)
+ \ldots\right)\right)
\nn
\ee
}

\noindent
When this operator acts on particular $\chi_{_R}$ only a few terms in the last two
brackets contribute, providing a polynomial in $z_{1}^{-1}$ and $z_2^{-1}$ of
the common degree $|R|$.
For example, the action on $\chi_{[1,1]}={\rm Mac}_{_{[1,1]}}$ gives
{\footnotesize
\be
\left(1+ \{q\}\left(\frac{q}{z_1}\hat\chi_{_{[1]}}
+ \frac{q^2}{z_1^2}\left(q\,\hat\chi_{_{[2]}}-\frac{1}{q}\,\hat\chi_{_{[1,1]}}\right)
+ \ldots\right)\right)
\left(\chi_{_{[1,1]}}+ \{q\}\left(\frac{q}{z_2}\chi_{_{[1]}}
+ \frac{q^2}{z_2^2}\cdot\Big(-\frac{1}{q}\Big) \right)\right) =
\nn \\
= \chi_{_{[1,1]}}+ \{q\}\left(\frac{q}{z_2}\chi_{_{[1]}}
+ \frac{q^2}{z_2^2}\cdot \Big(-\frac{1}{q}\Big) \right) +
 \frac{q\{q\}}{z_1} \left(\chi_{_{[1]}}+  \frac{q\{q\}}{z_2}\right)
+ \frac{q^2\{q\}}{z_1^2}\cdot \Big(-\frac{1}{q}\Big)
= \nn \\
= \chi_{_{[1,1]}} + q\{q\}\chi_{_{[1]}}\left(\frac{1}{z_1}+\frac{1}{z_2}\right)
- q\{q\}\left(\frac{1}{z_1^2} +\frac{1}{z_2^2}\right) +\frac{q^2\{q\}^2}{z_1z_2}
\nn
\ee
}

\noindent
Now we need to multiply by the first two brackets, but pick up only the terms
of total grading zero, since they will be selected by the contour integrals
over $z_1$ and $z_2$:
{\footnotesize
\be
:\hat V_1(z_1)\hat V_1(z_2): \,\chi_{_{[1,1]}} \ \longrightarrow \
\left( 1 + \frac{\{t\}}{t}(z_1+z_2)\chi_{_{[1]}}
 + \frac{\{t\}^2}{t^2}(z_1z_2)\chi_{_{[1]}}^2
 + \frac{\{t\}}{t^2}(z_1^2+z_2^2)
 \left(t\,\chi_{_{[2]}}-\frac{1}{t}\,\chi_{_{[1,1]}}\right)\right)\cdot
 \nn \\
\cdot \left(
\chi_{_{[1,1]}} + q\{q\}\chi_{_{[1]}}\left(\frac{1}{z_1}+\frac{1}{z_2}\right)
-q\{q\}\left(\frac{1}{z_1^2} +\frac{1}{z_2^2}\right) +\frac{q^2\{q\}^2}{z_1z_2}
\right) \ \longrightarrow \
\nn \\
\chi_{_{[1,1]}} +
\frac{q\{q\}\{t\}}{t}\frac{(z_1+z_2)^2}{z_1z_2}\chi_{_{[1]}}^2
- \frac{q\{q\}\{t\}^2}{t^2}\frac{z_1^2-q\{q\}z_1z_2+z_2^2}{z_1z_2}\chi_{_{[1]}}^2
- \frac{q\{q\}\{t\}}{t^2}\frac{(z_1^2-q\{q\}z_1z_2+z_2^2)(z_1^2+z_2^2)}{z_1^2z_2^2}
 \left(t\,\chi_{_{[2]}}-\frac{1}{t}\,\chi_{_{[1,1]}}\right)
\nn
\ee
}

\noindent
It remains to substitute the integrals
\be\label{int}
\frac{\{t^2\}}{\{t\}^2}
\oint\frac{dz_1}{z_1}\oint\frac{dz_2}{z_2}\frac{(z_1-z_2)^2}{(z_1-t^2z_2)(z_1-t^{-2}z_2)}
\left(\frac{z_1}{z_2}\right)^m
= \ \left\{\begin{array}{ccc}   \{t^{2m}\} & {\rm for} & m>0 \\
\frac{\{t^2\}}{\{t\}^2} & {\rm for} & m=0 \\
0 &{\rm for} & m< 0 \end{array} \right.
\ee
to get
{\footnotesize
\be
\frac{\{t^2\}}{\{t\}^2}
\oint\frac{dz_1}{z_1}\oint\frac{dz_2}{z_2}\frac{(z_1-z_2)^2}{(z_1-t^2z_2)(z_1-t^{-2}z_2)}
:\hat V_1(z_1)\hat V_1(z_2): \,\chi_{_{[1,1]}}
= \frac{\{t^2\}}{\{t\}^2}\cdot \chi_{_{[1,1]}} +
\frac{q\{q\}\{t\}}{t}\Big(\{t^2\} + 2\cdot\frac{\{t^2\}}{\{t\}^2} + 0\Big)  \chi_{_{[1]}}^2
+ \nn \\
- \frac{q \{q\}\{t\}^2}{t^2}\Big( \{t^2\} -q\{q\}\frac{\{t^2\}}{\{t\}^2}  + 0\Big)\chi_{_{[1]}}^2
- \frac{q \{q\}\{t\}}{t^2}\Big( \{t^4\} - q\{q\}\{t^2\} + 2\frac{\{t^2\}}{\{t\}^2} + 0 + 0\Big)
 \left(t\,\chi_{_{[2]}}-\frac{1}{t}\,\chi_{_{[1,1]}}\right)
\nn
\ee
}
This answer contains both $\chi_{_{[2]}}$ and $\chi_{_{[1,1]}}$,
thus $\chi_{_{[1,1]}}={\rm Mac}_{_{1,1]}}$ is {\it not} an  eigenfunction of
integrated $:\hat V_1(z_1)\hat V_1(z_2):$
This is cured by adding $\oint \frac{dz}{z} \,\hat V_2(z)$:
{\footnotesize
\be
\oint \frac{dz}{z} \,\hat V_2(z) \,\chi_{_{[1,1]}} =
\left\{1 +  \{q\}\{t^2\}\frac{q}{t}\Big(t+\frac{1}{t}\Big) \chi_{_{[1]}} \hat\chi_{{[1]}}
 + \{q\}\{t^2\}\Big(\frac{q}{t}\Big)^2\left(t\chi_{_{[2]}} -\frac{1}{t} \chi_{_{[1,1]}}\right)
\left(\Big(q[3]_t -\frac{1}{q}\Big)\hat\chi_{_{[2]}}
- \Big(\frac{[3]_t}{q}-{q}\Big)\hat\chi_{_{[1,1]}}\right)+\ldots \right\}\chi_{_{[1,1]}}
= \nn
\ee
}
\vspace{-0.4cm}
\be
= \chi_{_{[1,1]}} + \{q\}\{t^2\}\frac{q}{t}\Big(t+\frac{1}{t}\Big)  \chi_{_{[1]}}^2
+ \{q\}\{t^2\}\Big(\frac{q}{t}\Big)^2 \Big(q-\frac{[3]_t}{q}\Big)
\left(t^2\chi_{_{[2]}}-\frac{1}{t^2} \chi_{_{[1,1]}}\right)
\ee
Together they provide an answer, proportional to $\chi_{_{[1,1]}}={\rm Mac}_{_{1,1]}}$:
\be
\hat H_2\, \chi_{_{[1,1]}}
= -\frac{2\Big(q^4(t^8-t^6-t^4+t^2)+q^2(t^6+t^4 -t^2-1)+1\Big)}{t^9\{t\}}\cdot \chi_{_{[1,1]}}
\ee

\paragraph{An example: Hamiltonians ${\cal H}_Q$ at the first three levels.}

Here we consider the first three levels in order to illustrate the construction of the Hamiltonians ${\cal H}_Q$.
Note that (\ref{condforxi}) are formulated entirely in terms of skew {\it Schur} polynomials,
what makes the calculations easy (once the Kostka-Kerov matrix is available from \cite{knotebook}).

\begin{itemize}

\item{\bf Level 1.} Here $R=1$, $\Ker_{[1]}=p_1$ and $\boxed{\xi_{[1],[1]}=E_{[1]}}$

\item{\bf Level 2.} This example is already informative.
We have two Kerov functions $\Ker_{[1,1]}=\frac{p_2+p_1^2}{2}=\chi_{[1,1]}$ and
$\Ker_{[2]} =\chi_{[2]}+\frac{g_2-g_1^2}{g_2+g_1^2}\cdot\chi_{[1,1]}
= \frac{-g_1^2p_2+g_2p_1^2}{g_2+g_1^2}$.
Thus the two equations in (\ref{condforxi}) are:
\be
 (\hat{\cal H}-E_{[1,1]})\Ker_{[1,1]} =
 \xi_{_{[1],[1]}}\chi_{_{[1]}}^2 + \xi_{_{[2],[1,1]}}\chi_{_{[2]}}
+(\xi_{_{[1,1]}}-E_{[1,1]})\chi_{_{[1,1],[1,1]}} = 0 \nn \\
\nn \\
 (\hat{\cal H}-E_{[2]})\Ker_{[2]} =
 \left(\xi_{_{[1],[1]}}\chi_{_{[1]}}^2 + (\xi_{_{[2],[2]}}-E_{[2]})\chi_{_{[2]}}
+\xi_{_{[1,1],[2]}}\chi_{_{[1,1]}}\right)
+ \nn \\
+\frac{g_2-g_1^2}{g_2+g_1^2}\left(
 \xi_{_{[1],[1]}}\chi_{_{[1]}}^2 + \xi_{_{[2],[1,1]}}\chi_{_{[2]}}
+(\xi_{_{[1,1],[1,1]}}-E_{[2]})\chi_{_{[1,1]}}\right) = 0
\ee
We used here the fact that $\chi_{_{[2]/[1]}}=\chi_{_{[1,1]/[1]}}$
and can further use $\chi_{_{[1]}}^2=\chi_{_{[2]}}+\chi_{_{[1,1]}}$.
Moreover, the first equation implies that, in the last line, we can substitute the
bracket for just $(E_{[1,1]}-E_{[2]})\chi_{_{[1,1]}}$, i.e. a non-trivial $g$-dependence
appears only in the coefficient of $\chi_{_{[1,1],[1,1]}}$, not of $\chi_{_{[2]}}$.
In other words, we get a system
\be
\begin{array}{ccc}
\xi_{_{[1,1],[1,1]}} = E_{[1,1]}-\xi_{_{[1],[1]}} = E_{[1,1]}-E_{[1]}
&&
\xi_{_{[2],[1,1]}}=-\xi_{{[1],[1]}}=-E_{[1]}
 \\
\xi_{_{[1,1],[2]}} =  \frac{g_2-g_1^2}{g_2+g_1^2}(E_{[2]}-E_{[1,1]})  - E_{[1]}
&&
\xi_{_{[2],[2]}} = E_{[2]}-\xi_{_{[1],[1]}} = E_{[2]}-E_{[1]}
\end{array}
\ee
and
\vspace{-0.3cm}
\be
\xi =
\left(
 \begin{array}{ccc}

&\\
E_{[1,1]}-E_{[1]} && -E_{[1]} \\
&\\
 \frac{g_2-g_1^2}{g_2+g_1^2}(E_{[2]}-E_{[1,1]})  - E_{[1]} &\ \ & E_{[2]}-E_{[1]} \\
&\\
 \end{array}
\right)
\label{xilev2}
\ee
Thus, at this level, we can
already collect all the terms proportional to $E_{[1]}$,
and reveal the structure of the simplest Hamiltonian (\ref{HamQ}):
\be
\boxed{
\phantom{.^{^{^{\int^5}}}}\!\!\!\!
\hat{\cal H}_{[1]}
=  \chi_{_{[1]}}\hat\chi_{_{[1]}} - (\chi_{_{[2]}}+\chi_{_{[1,1]}})
(\hat\chi_{_{[2]}}+\hat\chi_{_{[1,1]}}) + \ldots
=   \chi_{_{[1]}} \cdot\Big(\hat\chi_{_{[1]}} - \chi_{_{[1]}} \cdot
(\hat\chi_{_{[2]}}+\hat\chi_{_{[1,1]}}) + \ldots
}
\label{H[1]}
\ee
associated with the eigenvalue $E_{[1]}$:
it annihilates all $\chi_{_R}$ except for $\chi_{_{[1]}}$,
\be
\hat{\cal H}_{[1]}\chi_{_R} =  \chi_{_R}\cdot\delta_{R,[1]}
\label{propH[1]}
\ee
Also seen at level two are the two other Hamiltonians, but only the first terms can be defined:
\be
\boxed{
\begin{array}{c}
\hat{\cal H}_{[1,1]} =   \chi_{_{[1,1]}}
\Big(\hat\chi_{_{[1,1]}} - \frac{g_2-g_1^2}{g_2+g_1^2}\,\hat\chi_{_{[2]}}+\ldots\Big)
  =   \Ker_{[1,1]}\cdot
\Big(\hat\chi_{_{[1,1]}} - \frac{g_2-g_1^2}{g_2+g_1^2}\,\hat\chi_{_{[2]}}+\ldots\Big)
\\ \\
\hat{\cal H}_{[2]} =  \chi_{_{[2]}}
\hat\chi_{_{2}} + \frac{g_2-g_1^2}{g_2+g_1^2}\,\chi_{_{[1,1]}}\hat\chi_{_{[2]}}
+ \ldots
=  \Ker_{[2]}\cdot\hat\chi_{_{[2]}}+\ldots
\end{array}
}
\ee
Note that, while $\hat{\cal H}_{[2]}$ annihilates $\chi_{[1,1]}=\Ker_{[1,1]}$,
the other Hamiltonian $\hat{\cal H}_{[1,1]}$ annihilates not $\chi_{[2]}$,
but $\Ker_{[2]} \sim \chi_{_{[2]}} + \frac{g_2-g_1^2}{g_2+g_1^2}\,\chi_{_{[1,1]}}$.

\item{\bf Level 3.}
Now
\be
\Ker_{[1,1,1]}=\chi_{_{[1,1,1]}} \ \ \ \ \ \
\Ker_{[2,1]} = \chi_{_{[2,1]}} + \frac{2(g_3-g_1^3)}{2g_3+3g_2g_1+g_1^3}\,\chi_{_{[1,1,1]}}
\nn \\
\Ker_{[3]} = \chi_{_{[3]}} + \frac{2g_2(g_3-g_1^3)}{g_3g_2+3g_3g_1^2+2g_2g_1^3}\, \chi_{_{[2,1]}}
+ \frac{g_3g_2-3g_3g_1^2+2g_2g_1^3}{g_3g_2+3g_3g_1^2+2g_2g_1^3}\,  \chi_{_{[1,1,1]}}
\ee
From these formulas, one can calculate the level-$3$ block of the matrix $\xi_{_{X,Y}}$:

\bigskip

\centerline{
{\footnotesize
$
\left(\begin{array}{ccc}
E_{[1,1,1]}-E_{[1,1]}  &E_{[1]}-E_{[1,1]}
+\frac{(g_2-g_1^2)(E_{[1,1]}-E_{[2]}) }{g_2+g_1^2}
+
& E_{[1]}+\frac{(g_2-g_1^2)(E_{[1,1]}-E_{[2]}) }{g_2+g_1^2}
+ \frac{(2g_3-3g_2g_1+g_1^3)(E_{[1,1,1]}-E_{[2,1]})}{2g_3+3g_2g_1+g_1^3}
+
\\
& + \frac{2(g_3-g_1^3)(E_{[2,1]}-E_{[1,1,1]})}{2g_3+3g_2g_1+g_1^3}
& +\frac{(g_2g_3-3g_3g_1^2+2g_2g_1^3)(E_{[3]}-E_{[2,1]})}{g_2g_3+3g_3g_1^2+2g_2g_1^3}\\
&&\\
E_{[1]}-E_{[1,1]}
&E_{[2,1]}+\frac{2g_1^2(E_{[1]}-E_{[1,1]})+2g_2(E_{[1]}-E_{[2]})}{g_2+g_1^2}
& \frac{-(g_2-g_1^2)(E_{[1]}-E_{[1,1]})+2g_2(E_{[1]}-E_{[2]})}{g_2+g_1^2}
+\frac{2g_2(g_3-g_1^3)(E_{[3]}-E_{[2,1]})}{g_2g_3+3g_3g_1^2+2g_2g_1^3}\\
&&\\
E_{[1]} & E_{[1]}-E_{[2]} & E_{[3]}-E_{[2]}
\end{array}\right)
$
}}

\bigskip

\bigskip

\noindent
Now one can read off the level-three contributions to the Hamiltonians:
\be
\!\!\!\!\!\!\!\!\!\!\!\!\!\!\!\!\!\!\!\!\!\!\!\!\!\!\!\!\!\!
\hat{\cal H}_{[1]}
=   \chi_{_{[1]}}\hat\chi_{_{[1]}} - (\chi_{_{[1,1]}}+\chi_{_{[2]}})
(\hat\chi_{_{[1,1]}}+\hat\chi_{_{[2]}})
+ (\chi_{_{[2,1]}}+\chi_{_{[3]}})\hat\chi_{_{[1,1,1]}}
+ (\chi_{_{[1,1,1]}}+2\chi_{_{[2,1]}}+\chi_{_{[3]}})\hat\chi_{_{[2,1]}}
+ (\chi_{_{[1,1,1]}}+\chi_{_{[2,1]}})\hat\chi_{_{[3]}}
+ \ldots =
\nn  \\
= \Ker_{[1]}\cdot\Big(\hat\chi_{_{[1]}}
- \chi_{_{[1]}} \cdot (\hat\chi_{_{[2]}}+\hat\chi_{_{[1,1]}})
+ \chi_{_{[1,1 ]}}(\hat\chi_{_{[2,1]}}+ \hat\chi_{_{[3]}})
+  \chi_{_{[2 ]}}( \hat\chi_{_{[1,1,1]}} + \hat\chi_{_{[2,1]}})
+ \ldots\Big)
\ \ \ \ \ \ \ \ \ \ \ \ \
\nn
\ee
\be
\!\!\!\!\!\!\!\!\!\!\!\!\!\!\!
\hat{\cal H}_{[1,1]} =  \chi_{_{[1,1]}}
\Big(\hat\chi_{_{[1,1]}} - \frac{g_2-g_1^2}{g_2+g_1^2}\,\hat\chi_{_{[2]}} \Big)
-  (\chi_{_{[1,1,1]}}+\chi_{_{[2,1]}})
\left((\hat\chi_{_{ [1,1,1]}}+\hat\chi_{_{ [2,1]}})
+ \frac{g_2-g_1^2}{g_2+g_1^2}\,(\hat\chi_{_{[2,1]}}+\hat\chi_{_{[3]}})
\right) + \ldots = \nn \\
=\Ker_{[1,1]}\cdot \left(
\Big(\hat\chi_{_{[1,1]}} - \chi_{_{[1]}}(\hat\chi_{_{ [1,1,1]}}+\hat\chi_{_{ [2,1]}})
+\ldots\Big)
- \frac{g_2-g_1^2}{g_2+g_1^2}\cdot\Big(\underline{\hat\chi_{_{[2]}}
- \chi_{_{[1]}}(\hat\chi_{_{ [2,1]}}+\hat\chi_{_{[3]}})+\ldots}\Big)\right)
\nn \\
\label{H2andH11}  \\
\hat{\cal H}_{[2]} =  \left(\chi_{_{[2]}}
 + \frac{g_2-g_1^2}{g_2+g_1^2}\,\chi_{_{[1,1]}}\right)\hat\chi_{_{[2]}}
- \left(( \chi_{_{[2,1]}}+ \chi_{_{[3]}})+
\frac{g_2-g_1^2}{g_2+g_1^2}\,(\chi_{_{[1,1,1]}}+\chi_{_{[2,1]}})\right)
(\hat\chi_{_{[1,2]}}+\hat\chi_{_{[3]}}) +\ldots
= \nn \\
= \Ker_{[2]}\cdot \Big(\underline{\hat\chi_{_{[2]}}
- \chi_{_{[1]}}\cdot (\hat\chi_{_{[1,2]}}+\hat\chi_{_{[3]}})  + \ldots}\Big)
\nn
\ee

\be
\hat{\cal H}_{[1,1,1]} = \chi_{_{[1,1,1]}}\cdot\left(\hat\chi_{_{[1,1,1]}}
-\frac{ 2(g_3-g_1^3) }{2g_3+3g_2g_1+g_1^3}\,\hat\chi_{_{[2,1]}}
+ \frac{2g_3-3g_2g_1+g_1^3}{2g_3+3g_2g_1+g_1^3}
\, \hat \chi_{_{[3]}}\right)
   +\ldots
\nn \\
\hat{\cal H}_{[2,1]} = \chi_{_{[2,1]}}\hat\chi_{_{[2,1]}}
+ \frac{2(g_3-g_1^3) }{2g_3+3g_2g_1+g_1^3}\,\chi_{_{[1,1,1]}}\hat\chi_{_{[2,1]}}
- \frac{2g_2(g_3-g_1^3) }{g_2g_3+3g_3g_1^2+2g_2g_1^3}\,\chi_{_{[2,1]}} \cdot\hat\chi_{_{[3]}}
-\nn\\
- \left(\frac{2g_3-3g_2g_1+g_1^3 }{2g_3+3g_2g_1+g_1^3}
+\frac{g_2g_3-3g_3g_1^2+2g_2g_1^3 }{g_2g_3+3g_3g_1^2+2g_2g_1^3}\right)
\chi_{_{[1,1,1]}}\cdot\hat\chi_{_{[3]}} +\ldots
= \nn \\
=\Ker_{[2,1]} \cdot \left(\hat\chi_{_{[2,1]}}
- \frac{2g_2(g_3-g_1^3) }{g_2g_3+3g_3g_1^2+2g_2g_1^3}\,\hat\chi_{_{[3]}}
+ \ldots \right)
\nn \\
\hat{\cal H}_{[3]} =   \left(
\frac{g_2g_3-3g_3g_1^2+2g_2g_1^3 }{g_2g_3+3g_3g_1^2+2g_2g_1^3}\, \chi_{_{[1,1,1]}}
+ \frac{2g_2(g_3-g_1^3) }{g_2g_3+3g_3g_1^2+2g_2g_1^3} \, \chi_{_{[2,1]}}
+  \chi_{_{[3]}}\right)\cdot \hat\chi_{_{[3]}} +\ldots
= \Ker_{[3]}\cdot\hat\chi_{_{[3]}} +\ldots
\nn
\ee

\end{itemize}


\begin{thebibliography}{12}

\bibitem{AGT} L.~Alday, D.~Gaiotto, Y.~Tachikawa,
  Lett.Math.Phys.\ {\bf 91} (2010) 167-197, arXiv:0906.3219\\
  N.~Wyllard,
  JHEP {\bf 0911} (2009) 002, arXiv:0907.2189\\
  A.~Mironov, A.~Morozov, Nucl.Phys.\ {\bf B825} (2009) 1-37,
  arXiv:0908.2569

\bibitem{AF} H. Awata, B. Feigin, A. Hoshino, M. Kanai, J. Shiraishi, S. Yanagida, arXiv:1106.4088\\
Y.~Ohkubo, H.~Awata, H.~Fujino,
  arXiv:1512.08016

\bibitem{japsgenmac}
M. Fukuda, Y. Ohkubo, J. Shiraishi, arXiv:1903.05905

\bibitem{SF1}
A. Mironov, A. Morozov and Sh. Shakirov, JHEP {\bf 1102} (2011) 067 arXiv:1012.3137\\
V. Alba, V. Fateev, A. Litvinov, G. Tarnopolsky, Lett.Math.Phys. {\bf 98} (2011) 33-64, arXiv:1012.1312\\
A. Mironov, A. Morozov, S. Shakirov, A. Smirnov, Nucl. Phys. {\bf B855} (2012) 128, arXiv:1105.0948

\bibitem{genmac}
A.~Morozov, A.~Smirnov,
   Lett.Math.Phys.\ {\bf 104} (2014) 585, arXiv:1307.2576\\
S.~Mironov, An.~Morozov, Y.~Zenkevich,
JETP Lett.\  {\bf 99} (2014) 109, arXiv:1312.5732 \\
Y.~Ohkubo, arXiv:1404.5401 \\
Y.~Kononov and A.~Morozov,  Eur.Phys.J. {\bf C76} (2016)  424,  arXiv:1607.00615     \\
Y.~Zenkevich,  arXiv:1612.09570

\bibitem{SF2}
M. Aganagic, Sh. Shakirov, arXiv:1105.5117; arXiv:1210.2733\\
P. Dunin-Barkowski, A. Mironov, A. Morozov, A. Sleptsov,
A. Smirnov, JHEP {\bf 03} (2013) 021,
arXiv:1106.4305\\
I. Cherednik, arXiv:1111.6195\\
A. Mironov, A. Morozov, An. Morozov,
JHEP 1203 (2012) 034, arXiv:1112.2654\\
A. Mironov, A. Morozov, Sh. Shakirov, A. Sleptsov, JHEP {\bf 2012} (2012) 70,
arXiv:1201.3339\\
A. Mironov, A. Morozov, Sh. Shakirov, J. Phys. A: Math. Theor. {\bf 45} (2012) 355202, arXiv:1203.0667

\bibitem{MMMI} A. Mironov, A. Morozov, An. Morozov,
in: {\it Strings, Gauge Fields, and the Geometry Behind: The Legacy of Maximilian Kreuzer},
eds: A.Rebhan, L.Katzarkov, J.Knapp, R.Rashkov, E.Scheidegger,
World Scietific,   2013 pp.101-118
arXiv:1112.5754

\bibitem{CS} S.-S. Chern, J. Simons,
Ann.Math. {\bf 99} (1974) 48-69\\
E. Witten,
Comm.Math.Phys. {\bf 121} (1989)  351-399

\bibitem{knotpols} J.W. Alexander, 
Trans.Amer.Math.Soc. {\bf 30} (2) (1928) 275-306\\
V.F.R. Jones, 
Invent.Math. {\bf 72} (1983) 1
Bull.AMS {\bf 12} (1985) 103
Ann.Math. {\bf 126} (1987) 335\\
L. Kauffman, 
Topology, {\bf 26} (1987) 395\\
P. Freyd, D. Yetter, J. Hoste, W.B.R. Lickorish, K. Millet,
A. Ocneanu,
Bull. AMS. {\bf 12} (1985) 239\\
J.H. Przytycki, K.P. Traczyk, 
Kobe J. Math. {\bf 4} (1987) 115-139\\
J.H. Conway, 
Algebraic Properties,
In: John Leech (ed.), {\sl Computational Problems in Abstract Algebra}, Proc.
Conf.
Oxford, 1967, Pergamon Press, Oxford-New York, 329-358, 1970

\bibitem{Mac} I.G. Macdonald, {\sl Symmetric functions and Hall polynomials},
Second Edition, Oxford University Press, 1995

\bibitem{AFS} H.~Awata, B.~Feigin, J.~Shiraishi,
  JHEP {\bf 1203} (2012) 041, arXiv:1112.6074

\bibitem{DIMMac}
H.~Awata, H.~Kanno, T.~Matsumoto, A.~Mironov, A.~Morozov, A.~Morozov, Y.~Ohkubo, Y.~Zenkevich,
  JHEP {\bf 1607} (2016)  103,
arXiv:1604.08366

\bibitem{AR}
H.~Awata, H.~Kanno, A.~Mironov, A.~Morozov, A.~Morozov, Y.~Ohkubo, Y.~Zenkevich,
JHEP {\bf 1610} (2016) 047, arXiv:1608.05351

\bibitem{DIM} J. Ding, K. Iohara, 
Lett.Math.Phys. {\bf 41} (1997) 181--193, q-alg/9608002\\
K. Miki, J. Math. Phys. {\bf 48} (2007) 123520

\bibitem{MacMahon} Y. Zenkevich, arXiv:1712.10300 \\
A. Morozov, Phys.Lett. {\bf B785} (2018) 175-183,  arXiv:1808.01059 \!;  arXiv:1810.00395

\bibitem{Kerov} S.V. Kerov, Func.An.and Apps. {\bf 25} (1991) 78-81

\bibitem{FZ} P.G.O. Freund, A.V. Zabrodin, Phys.Lett. {\bf B294} (1992) 347–353, hep-th/9208063

\bibitem{Cha}   T.H. Baker, {\sl Symmetric functions and infinite-dimensional algebras,} PhD Thesis, 1994, Tasmania\\
A.H. Bougourzi, L. Vinet, Letters in Mathematical Physics {\bf 39} (1997) 299–311, q-alg/9604021\\
A.A. Bytsenko, M. Chaichian, R.J. Szabo, A. Tureanu, arXiv:1308.2177\\
A.A. Bytsenko, M. Chaichian, R. Luna, J.Math.Phys. {\bf 58} (2017) 121701, arXiv:1707.01553

\bibitem{MMkerov} A.Mironov and A.Morozov,  arXiv:1811.01184

\bibitem{RHam} S.N.M. Ruijsenaars, H. Schneider,
Ann.Phys. (NY), {\bf 170} (1986) 370\\
S.N.M. Ruijsenaars, Comm.Math.Phys. {\bf 110} (1987) 191-213;
Comm.Math.Phys. {\bf 115} (1988) 127-165

\bibitem{MMgenmac} A. Mironov, A. Morozov, arXiv:1907.05410\\
A. Mironov, A. Morozov, Y. Zenkevich, to appear

\bibitem{MMN1} A.Mironov, A.Morozov, S.Natanzon,
Theor.Math.Phys. {\bf 166} (2011) 1-22,  arXiv:0904.4227;
Journal of Geometry and Physics, {\bf 62} (2012) 148-155, arXiv:1012.0433

\bibitem{GJ}  D. Goulden , D.M. Jackson, A. Vainshtein,
Ann. of Comb. {\bf 4} (2000) 27-46,
Brikh\"auser, math/9902125

\bibitem{KI} V. Ivanov, S. Kerov, 
Journal of Mathematical Sciences (Kluwer) {\bf 107} (2001) 4212-4230, math/0302203

\bibitem{MMNspin} A. Mironov, A. Morozov, S. Natanzon, arXiv:1904.11458

\bibitem{MmacW}
A. Morozov, Theor.Math.Phys. {\bf 200} (2019) 938-965,  arXiv:1810.00395

\bibitem{MMsumrules}  A. Mironov, A. Morozov, JHEP {\bf 08} (2018) 163, arXiv:1807.02409

\bibitem{triangle}
A.Morozov,   arXiv:1906.09971

\bibitem{Mcauchy}
A. Morozov, Eur.Phys.J. {\bf C79} (2019) no.1, 76, arXiv:1812.03853

\bibitem{MMkerov2} A. Mironov, A. Morozov, Nucl.Phys. {\bf B944} (2019) 114641,  arXiv:1903.00773

\bibitem{KM}  M. Kashiwara, T. Miwa, Proc. Japan cad. {\bf A57} (1981) 342-347

\bibitem{DJKMIII} E. Date, M. Jimbo, M. Kashiwara, T. Miwa, J. Phys. Soc. Jpn. {\bf 50} (1981) 3806-3812

\bibitem{knotebook} http://knotebook.org.s3-website-us-west-2.amazonaws.com/knotebook/KerMacData/kermac.htm

\end{thebibliography}
\end{document}